\documentclass[twocolumn,showpacs,preprintnumbers,amsmath,amssymb]{revtex4}

\usepackage{graphicx}
\usepackage{dcolumn}
\usepackage{bm}

\begin{document}

\title{Dynamics of non-equilibrium thermal entanglement}

\author{Ilya Sinaysky}
 \email{ilsinay@gmail.com}
\affiliation{
Quantum Research Group, School of Physics,
 University of KwaZulu-Natal, Durban, 4001, South Africa}

\author{Francesco Petruccione}
 \email{petruccione@ukzn.ac.za}
\affiliation{
Quantum Research Group, School of Physics and National Institute
for Theoretical Physics,
 University of KwaZulu-Natal, Durban, 4001, South Africa}

\author{Daniel Burgarth}
 \email{daniel.burgarth@maths.ox.ac.uk}
\affiliation{ Mathematical Institute, University of Oxford, 24-29
St Giles' Oxford OX1 3LB, UK}

\date{\today}

\begin{abstract}
The dynamics of a simple spin chain (2 spins) coupled to bosonic
baths at different temperatures is studied. The analytical
solution for the reduced density matrix of the system is found.
The dynamics and temperature dependence of spin-spin entanglement
is analyzed. It is shown that the system converges to a steady-state.
If the energy levels of the two spins are different, the steady-state
concurrence assumes its maximum at unequal bath temperatures.
It is found that a difference in local energy levels can make the steady-state
entanglement more stable against high temperatures.
\end{abstract}

\maketitle

\section{INTRODUCTION}

In describing real physical systems one should always take into
account the influence of the surroundings. The study of open
systems is particularly important for understanding processes in
quantum physics~\cite{toqs}. Whereas in most cases the
interaction with an environment destroys quantum correlations
within the system, it is well known that in some situations it
can also build up entanglement~\cite{braun} and in principle even
prepare complex entangled states~\cite{diehl}. The dynamics of
entanglement in open systems provides many interesting insights
into relaxation and transport situations, in particular if the
system dynamics involves many-body interactions (such as spin
chains, see~\cite{vedral} for a review). In order to understand
the role of the various parameters that compete in this setup, it
is useful to find exactly solvable models. Here we study the
dynamics of a model that was recently introduced by L.
Quiroga~\cite{Qui}. It consists of a simple spin chain in contact
with two reservoirs at different temperatures. In such a
non-equilibrium case most studies are restricted to the
steady-state to which the system converges in the limit
of long times~\cite{Qui,prosen,key-6,medi}. The dynamics for the
model in the zero-temperature limit was studied in~\cite{key-7}.
In the following, we study the dynamics of this model for generic
temperatures.

This paper is organized as follows. In Sec. II we describe the
model of a spin chain coupled to bosonic baths at different
temperatures as introduced in Ref.
\cite{Qui}. For completeness we follow \cite{Qui} in deriving the
master equation for the reduced density matrix in the Born-Markov
approximation. In Sec. III we present the analytical solution for
the system dynamics and show the convergence of the obtained
solution to the density matrix of the non-equilibrium steady
state solution. One should note that in \cite{Qui} this
steady-state was found only in the case when the energy levels of the spins are equal. Finally, in Sec.
IV we discuss the results and conclude.

\section{MODEL}

We consider the simplest spin chain consisting of two spins, with
each spin coupled to a separate bosonic bath. In the derivation of
the master equation we follow the formalism suggested in Ref.
\cite{Qui}. The total Hamiltonian is given by \begin{equation}
\hat{H}=\hat{H}_{S}+\hat{H}_{B1}+\hat{H}_{B2}+\hat{H}_{SB1}+\hat{H}_{SB2},\nonumber\end{equation}
where \[
\hat{H}_{S}=\frac{\epsilon_{1}}{2}\hat{\sigma}_{1}^{z}+\frac{\epsilon_{2}}{2}\hat{\sigma}_{2}^{z}+K(\hat{\sigma}_{1}^{+}\hat{\sigma}_{2}^{-}+\hat{\sigma}_{1}^{-}\hat{\sigma}_{2}^{+})\]
 is the Hamiltonian describing spin-to-spin interactions and $\hat{\sigma}_{i}^{z},\hat{\sigma}_{i}^{\pm}$
are the Pauli matrices. In this paper units are chosen such that
$k_B=\hbar=1$. The constants $\epsilon_{1}$ and $\epsilon_{2}$
denote the energy of spins 1 and 2, respectively, whereas $K$
denotes the strength of spin-spin interaction. We will see later
that the energy difference
$\Delta\epsilon=\epsilon_{1}-\epsilon_{2}$ has a crucial role in
determining the entanglement of the thermal state. We refer to
the case $\Delta\epsilon=0$ studied in \cite{Qui} as the
\emph{symmetric} case. Our study focuses on the non-symmetric
case $\Delta\epsilon\neq0$. The Hamiltonians of the reservoirs
for each spin $j=1,2$ are given by
$$
\hat{H}_{Bj}=\sum_n \omega_{n,j}\hat{b}^+_{n,j}\hat{b}_{n,j}.
$$
 The interaction between the spin subsystem and the bosonic baths is described by
\begin{equation}
\hat{H}_{SBj}=\hat{\sigma}_{j}^{+}\sum_{n}g_{n}^{(j)}\hat{b}_{n,j}+\hat{\sigma}_{j}^{-}\sum_{n}g_{n}^{(j)*}\hat{b}_{n,j}^{\dag}\equiv\sum_{\mu}\hat{V}_{j,\mu}\hat{f}_{j,\mu}.\nonumber\end{equation}
The operators $\hat{V}_{j,\mu}$ are chosen to satisfy
$[\hat{H}_{S},\hat{V}_{j,\mu}]=\omega_{j,\mu}\hat{V}_{j,\mu},$ and
the $\hat{f}_{j,\mu}$ act on bath degrees of freedom (this is
always possible; their explicit form will be given later on).
Physically, the index $\mu$ corresponds to \emph{transitions
}between eigenstates of the system induced by the bath. The whole
system (spin chain with reservoirs) is described by the Liouville
equation
\[ \frac{d}{dt}\hat{\alpha}=-i[\hat{H},\hat{\alpha}].\] We assume
that the evolution of the dynamical subsystem (coupled spins) does
not influence the state of the environment (bosonic reservoirs) so
that the density operator of the whole system $\hat{\alpha}(t)$
can be written as: \[
\hat{\alpha}(t)=\hat{\rho}(t)\hat{B}_{1}(0)\hat{B}_{2}(0)\]
 (irreversibility hypothesis), where each bosonic bath is described
by a canonical density matrix
$\hat{B}_{j}=e^{-\beta_{j}\hat{H}_{Bj}}/\textrm{tr}[e^{-\beta_{j}\hat{H}_{Bj}}]$
and $\hat{\rho}(t)$ denotes the reduced density matrix of the spin
chain.

In Born-Markov approximation the equation for the evolution of the
reduced density matrix \cite{goldman} is: \[
\frac{d\hat{\rho}}{dt}=-i[\hat{H}_{S},\hat{\rho}]+{\cal
L}_{1}(\hat{\rho})+{\cal L}_{2}(\hat{\rho})\]
 with dissipators

\begin{eqnarray*}
\mathcal{L}_{j}(\hat{\rho})\equiv\sum_{\mu,\nu}J_{\mu,\nu}^{(j)}(\omega_{j,\nu})\{[\hat{V}_{j,\mu},[\hat{V}_{j,\nu}^{\dag},\hat{\rho}]]-\\
-(1-e^{\beta_{j}\omega_{j,\nu}})[\hat{V}_{j,\mu},\hat{V}_{j,\nu}^{\dag}\hat{\rho}]\}\end{eqnarray*}
 and where the spectral density is given by

\[
J_{\mu,\nu}^{(j)}(\omega_{j,\nu})=\int_{0}^{\infty}dse^{i\omega_{j,\nu}s}\langle
e^{-is\hat{B}_{j}}\hat{f}_{j,\nu}^{\dag}e^{is\hat{B}_{j}}\hat{f}_{j,\mu}\rangle_{j}.\]
To find a solution we go to the basis of the eigenvectors
$|\lambda_{i}\rangle$ with eigenvalues $\lambda_{i}$ of the
Hamiltonian $\hat{H}_{S},$
\[
|\lambda_{1}\rangle=|0,0\rangle,\quad\lambda_{1}=-\frac{\epsilon_1+\epsilon_2}{2},\]
 \[
|\lambda_{2}\rangle=|1,1\rangle,\quad\lambda_{2}=\frac{\epsilon_1+\epsilon_2}{2},\]
 \[
|\lambda_{3}\rangle={\rm cos}(\theta/2)|1,0\rangle+{\rm
sin}(\theta/2)|0,1\rangle,\quad\lambda_{3}=\kappa,\]
 \[
|\lambda_{4}\rangle=-{\rm sin}(\theta/2)|1,0\rangle+{\rm
cos}(\theta/2)|0,1\rangle,\quad\lambda_{4}=-\kappa,\]
 where $\kappa\equiv\sqrt{K^{2}+\frac{(\Delta\epsilon)^{2}}{4}}$
and ${\rm tan}\theta\equiv2K/(\Delta\epsilon)$. In this
representation the dissipative operator ${\cal L}_{i}(\hat{\rho})$
becomes \[ {\cal
L}_{j}(\hat{\rho})=\sum_{\mu=1}^{2}J^{(j)}(-\omega_{\mu})(2\hat{V}_{j,\mu}\hat{\rho}\hat{V}_{j,\mu}^{\dag}-\{\hat{\rho},\hat{V}_{j,\mu}^{\dag}\hat{V}_{j,\mu}\}_+)+\]
 \[
J^{(j)}(\omega_{\mu})(2\hat{V}_{j,\mu}^{\dag}\hat{\rho}\hat{V}_{j,\mu}-\{\hat{\rho},\hat{V}_{j,\mu}\hat{V}_{j,\mu}^{\dag}\}_+),\]
 with transition frequencies\[
\omega_{1}=\lambda_{2}-\lambda_{3},\]
\[\omega_{2}=\lambda_{2}+\lambda_{3}\]
and transition operators \[
\hat{V}_{1,1}={\rm
cos}(\theta/2)(|\lambda_{1}\rangle\langle\lambda_{3}|+|\lambda_{4}\rangle\langle\lambda_{2}|),\]
 \[ \hat{V}_{1,2}={\rm
sin}(\theta/2)(|\lambda_{3}\rangle\langle\lambda_{2}|-|\lambda_{1}\rangle\langle\lambda_{4}|),\]
 \[\hat{V}_{2,1}={\rm
sin}(\theta/2)(|\lambda_{1}\rangle\langle\lambda_{3}|-|\lambda_{4}\rangle\langle\lambda_{2}|)
,\]
 \[\hat{V}_{2,2}={\rm
cos}(\theta/2)(|\lambda_{3}\rangle\langle\lambda_{2}|+|\lambda_{1}\rangle\langle\lambda_{4}|)
.\]
In this paper we consider the bosonic bath as an infinite set of
harmonic oscillators, so the spectral density has the form
$J^{(j)}(\omega_{\mu})=\gamma_{j}(\omega_{\mu})n_{j}(\omega_{\mu})$,
where $n_{j}(\omega_{\mu})=(e^{\beta_{j}\omega_{\mu}}-1)^{-1}$ and
$J^{(j)}(-\omega_{\mu})=e^{\beta_{j}\omega_{\mu}}J^{(j)}(\omega_{\mu})$.
For simplicity we choose the coupling constant to be frequency
independent $\gamma_{1}(\omega)=\gamma_1$ and
$\gamma_{2}(\omega)=\gamma_2.$ In  the basis $|\lambda_i\rangle$ the equation for the
diagonal elements of the reduced density matrix is given by \[
\frac{d}{dt}\left(\begin{array}{c}
\rho_{11}(t)\\
\rho_{22}(t)\\
\rho_{33}(t)\\
\rho_{44}(t)\end{array}\right)=B\left(\begin{array}{c}
\rho_{11}(t)\\
\rho_{22}(t)\\
\rho_{33}(t)\\
\rho_{44}(t)\end{array}\right),\] where $B$ is a $4\times4$ matrix
with constant coefficients.
 The time-dependence for the non-diagonal elements has the following
form

\begin{equation*}
\rho_{i,j}(t)=e^{ts_{i,j}}\rho_{i,j}(0),\end{equation*} where
$s_{i,j}$ is a complex number. For the initial state of system in
the computational basis $\left\{
|00\rangle,|01\rangle,|10\rangle,|11\rangle\right\} $ we
choose\begin{eqnarray*}
\lefteqn{\hat{\rho}(0)=p_{0}|00\rangle\langle00|+p_{1}|01\rangle\langle01|+p_{2}|10\rangle\langle10|}\\
 & +(1-p_{0}-p_{1}-p_{2})|11\rangle\langle11|+c_{12}|01\rangle\langle10|+c_{12}^{*}|10\rangle\langle01|.\end{eqnarray*}

\section{EXACT SOLUTION}
The analytical solution in the basis of eigenvectors
$|\lambda_{i}\rangle$ is given by:

\[
\rho_{ii}(t)=\frac{1}{X_{1}Y_{2}}\left(\begin{array}{cccc}
a_{11} & a_{12} & a_{13} & a_{14}\\
a_{21} & a_{22} & a_{23} & a_{24}\\
a_{31} & a_{32} & a_{33} & a_{34}\\
a_{41} & a_{42} & a_{43} &
a_{44}\end{array}\right)\left(\begin{array}{c}
\rho_{11}(0)\\
\rho_{22}(0)\\
\rho_{33}(0)\\
\rho_{44}(0)\end{array}\right),\]
 where the coefficients $a_{ij}$ are given by:
 \[a_{11}=(X_{1}^{+}+X_{1}^{-}e^{-tX_{1}})(Y_{2}^{+}+Y_{2}^{-}e^{-tY_{2}}),\]
 \[a_{12}=(1-e^{-tX_{1}})(1-e^{-tY_{2}})X_{1}^{+}Y_{2}^{+},\]
 \[a_{13}=(1-e^{-tX_{1}})X_{1}^{+}(Y_{2}^{+}+Y_{2}^{-}e^{-tY_{2}}),\]
 \[a_{14}=(X_{1}^{+}+X_{1}^{-}e^{-tX_{1}})(1-e^{-tY_{2}})Y_{2}^{+},\]
 \[a_{21}=(1-e^{-tX_{1}})(1-e^{-tY_{2}})X_{1}^{-}Y_{2}^{-},\]
 \[a_{22}=(X_{1}^{-}+X_{1}^{+}e^{-tX_{1}})(Y_{2}^{-}+Y_{2}^{+}e^{-tY_{2}}),\]
 \[a_{23}=(X_{1}^{-}+X_{1}^{+}e^{-tX_{1}})(1-e^{-tY_{2}})Y_{2}^{-},\]
 \[a_{24}=(1-e^{-tX_{1}})X_{1}^{-}(Y_{2}^{-}+Y_{2}^{+}e^{-tY_{2}}),\]
 \[a_{31}=(1-e^{-tX_{1}})X_{1}^{-}(Y_{2}^{+}+Y_{2}^{-}e^{-tY_{2}}),\]
 \[a_{32}=(X_{1}^{-}+X_{1}^{+}e^{-tX_{1}})(1-e^{-tY_{2}})Y_{2}^{+},\]
 \[a_{33}=(X_{1}^{-}+X_{1}^{+}e^{-tX_{1}})(Y_{2}^{+}+Y_{2}^{-}e^{-tY_{2}}),\]
 \[a_{34}=(1-e^{-tX_{1}})(1-e^{-tY_{2}})X_{1}^{-}Y_{2}^{+},\]
 \[a_{41}=(X_{1}^{+}+X_{1}^{-}e^{-tX_{1}})(1-e^{-tY_{2}})Y_{2}^{-},\]
 \[a_{42}=(1-e^{-tX_{1}})X_{1}^{+}(Y_{2}^{-}+Y_{2}^{+}e^{-tY_{2}}),\]
 \[a_{43}=(1-e^{-tX_{1}})(1-e^{-tY_{2}})X_{1}^{+}Y_{2}^{-},\]
 \[a_{44}=(X_{1}^{+}+X_{1}^{-}e^{-tX_{1}})(Y_{2}^{-}+Y_{2}^{+}e^{-tY_{2}}).\]
Taking into account the initial conditions, the non-vanishing
non-diagonal elements are: \[
\rho_{34}(t)=e^{-i2t\lambda_{3}-\frac{t(X_{1}+Y_{2})}{2}}\rho_{34}(0),\]
 \[
\rho_{43}(t)=\bar{\rho}_{34}=e^{i2t\lambda_{3}-\frac{t(X_{1}+Y_{2})}{2}}\rho_{43}(0).\]
 In the present solution we have introduced some constants: \[
X_{i}=X_{i}^{+}+X_{i}^{-},\]
\[Y_{i}=Y_{i}^{+}+Y_{i}^{-},\]
$$ X_{i}^{\mp}=2{\rm cos}^2(\theta/2)J^{(1)}(\pm\omega_{i})+2{\rm sin}^2(\theta/2)J^{(2)}(\pm\omega_{i})$$
$$ Y_{i}^{\mp}=2{\rm sin}^2(\theta/2)J^{(1)}(\pm\omega_{i})+2{\rm cos}^2(\theta/2)J^{(2)}(\pm\omega_{i})$$
or
\begin{eqnarray*}
X_{i}^{\mp} &=& (J^{(1)}(\pm\omega_{i})+J^{(2)}(\pm\omega_{i}))\\
            & &+\frac{\Delta\epsilon}{\sqrt{4K^2+(\Delta\epsilon)^2}}(J^{(1)}(\pm\omega_{i})-J^{(2)}(\pm\omega_{i}))
\end{eqnarray*}
\begin{eqnarray*}
Y_{i}^{\mp} &=&(J^{(1)}(\pm\omega_{i})+J^{(2)}(\pm\omega_{i}))\\
            & &-\frac{\Delta\epsilon}{\sqrt{4K^2+(\Delta\epsilon)^2}}(J^{(1)}(\pm\omega_{i})-J^{(2)}(\pm\omega_{i})).\end{eqnarray*}
One can easily see that this solution converges with increasing time
to a diagonal density matrix which does not depend on the initial
conditions: \[
\lim_{t\rightarrow\infty}\rho_{ii}(t)=\frac{1}{X_{1}Y_{2}}\left(\begin{array}{c}
X_{1}^{+}Y_{2}^{+}\\
X_{1}^{-}Y_{2}^{-}\\
X_{1}^{-}Y_{2}^{+}\\
X_{1}^{+}Y_{2}^{-}\end{array}\right),\]

\[
\lim_{t\rightarrow\infty}\rho_{34}(t)=0.\] In the symmetric case ($\Delta\epsilon=0$) the above limit in
reproduces the result obtained by Quiroga in \cite{Qui}.
In order to quantify the entanglement between the spins we consider the concurrence~\cite{woot}.
In the steady-state
$(t\rightarrow\infty)$ it is given by\begin{eqnarray*}
C_{\infty}=\frac{2}{X_1Y_2}\textrm{Max}(0,\frac{{\rm
sin}\theta}{2}|X_1^+Y_2^--X_1^-Y_2^+|\\-\sqrt{X_1^-X_1^+Y_2^-Y_2^+})
.\end{eqnarray*}

\begin{figure}
\includegraphics[scale=0.4]{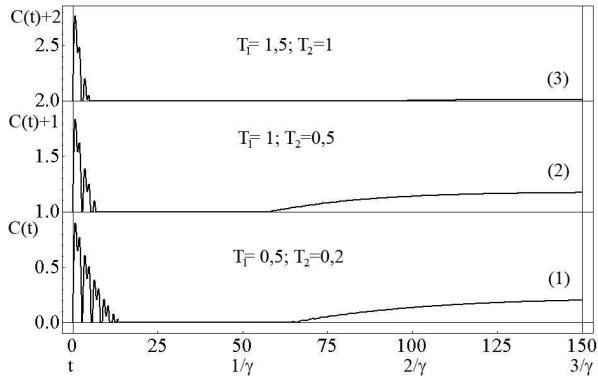}
\caption{Dynamics of the concurrence $C(t)$ for the initial
reduced density matrix $\hat{\rho}_0=|1,0\rangle\langle1,0|$. The
parameters of the model chosen to be $\gamma_1=\gamma_2=0.02$,
$\epsilon_1=2$, $\epsilon_2=1$, $K=1$ for different temperatures
of baths: curve (1) corresponds to $T_1=0,5;T_2=0,2$; curve (2)
 $T_1=1;T_2=0,5$; curve (3) $T_1=1,5;T_2=1$.}
\end{figure}

\begin{figure}
\includegraphics[scale=0.4]{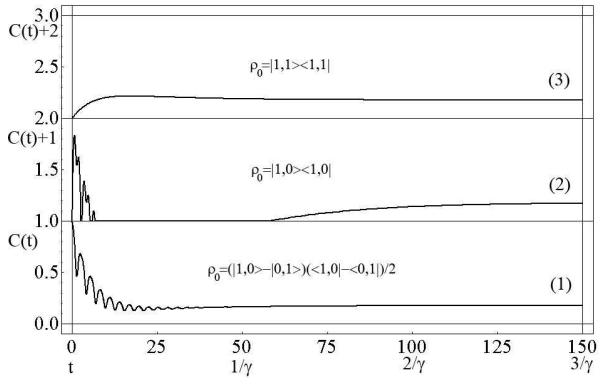}
\caption{Dynamics of the concurrence $C(t)$ for different initial
states of the reduced density matrix of qubits; $T_1=1$,
$T_2=0.5$, $\gamma=0.02$, $\epsilon_1=2$, $\epsilon_2=1$, $K=1$.
The curve (1) corresponds to
$\hat{\rho}_0=(|1,0\rangle-|0,1\rangle)(\langle1,0|-\langle0,1|)/2$;
curve (2) corresponds to $\hat{\rho}_0=|1,0\rangle\langle1,0|$;
curve (3) corresponds to $\hat{\rho}_0=|1,1\rangle\langle1,1|$.}
\end{figure}

\begin{figure}
\includegraphics[scale=0.4]{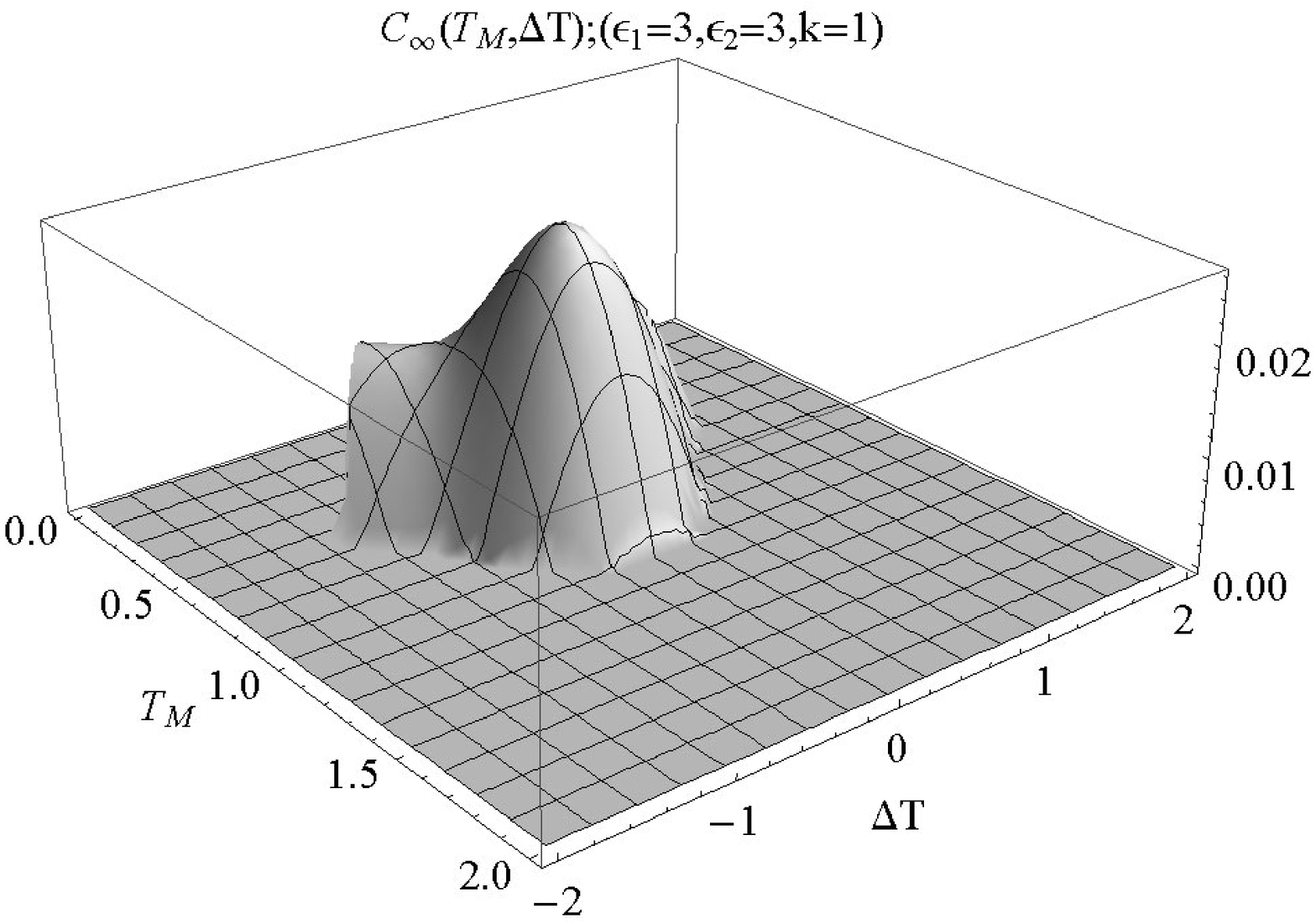}
\caption{Steady-state concurrence $C_{\infty}(T_M,\Delta T)$ as a
function of the mean bath temperature $T_M=(T_1+T_2)/2$ and
temperature difference $\Delta T=T_1-T_2$ in the symmetric case
$\epsilon_1=\epsilon_2=3$ with $K=1$.}
\end{figure}

\begin{figure}
\includegraphics[scale=0.4]{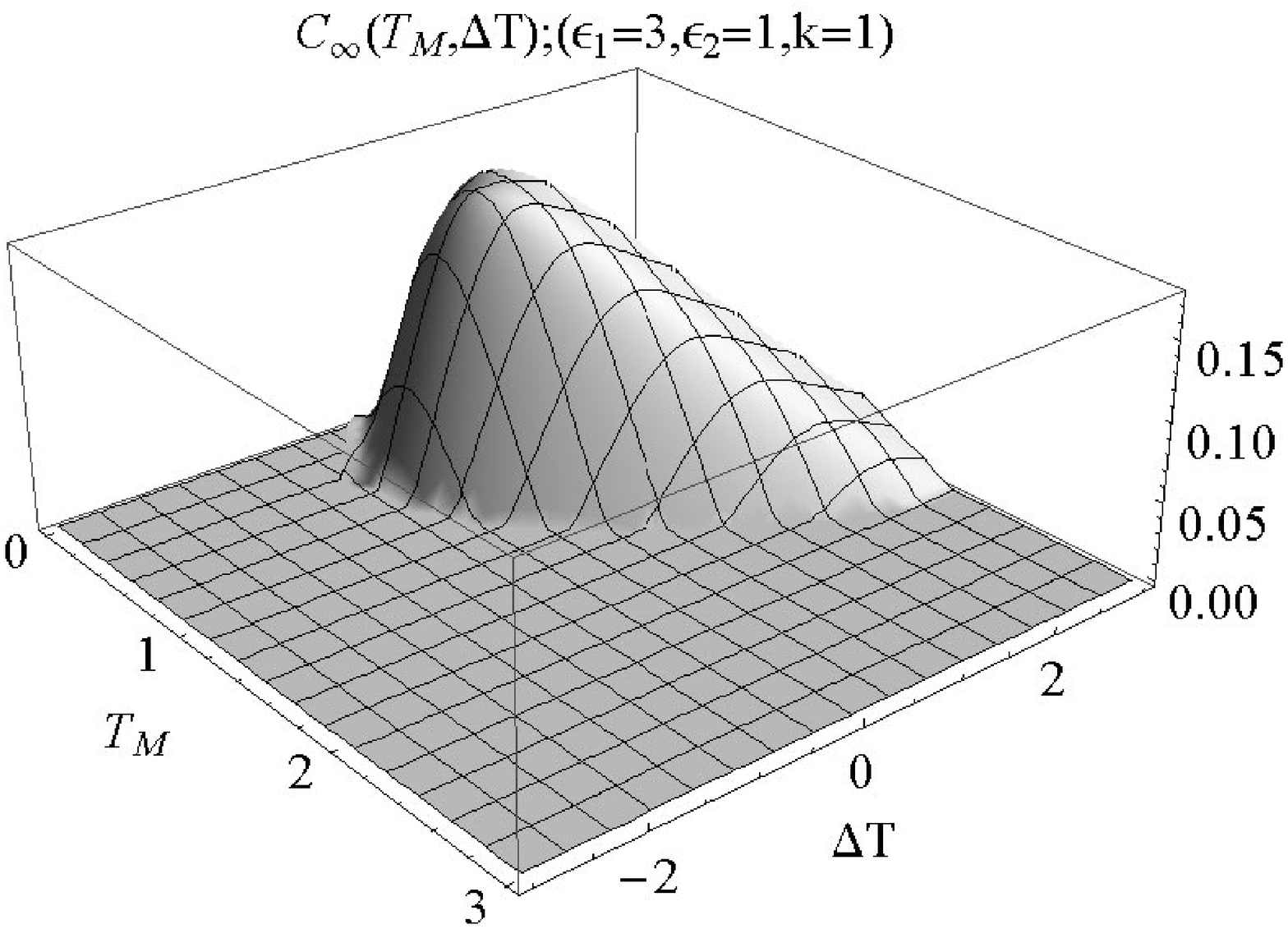}
\caption{Steady-state concurrence $C_{\infty}(T_M,\Delta T)$ as a
function of the mean bath temperature $T_M=(T_1+T_2)/2$ and the
temperature difference $\Delta T=T_1-T_2$ in the case
$\epsilon_1=3$, $\epsilon_2=1$, $K=1$.}
\end{figure}

\begin{figure}
\includegraphics[scale=0.4]{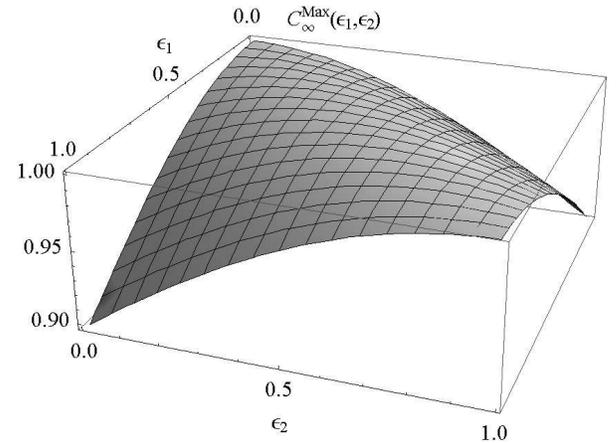}
\caption{Maximally possible steady-state concurrence
$C_{\infty}^{\mathrm{Max}}(\epsilon_1,\epsilon_2)$ in the strong
coupling case $\epsilon_1,\epsilon_2<K$ with $K=1$.}
\end{figure}

\begin{figure}
\includegraphics[scale=0.4]{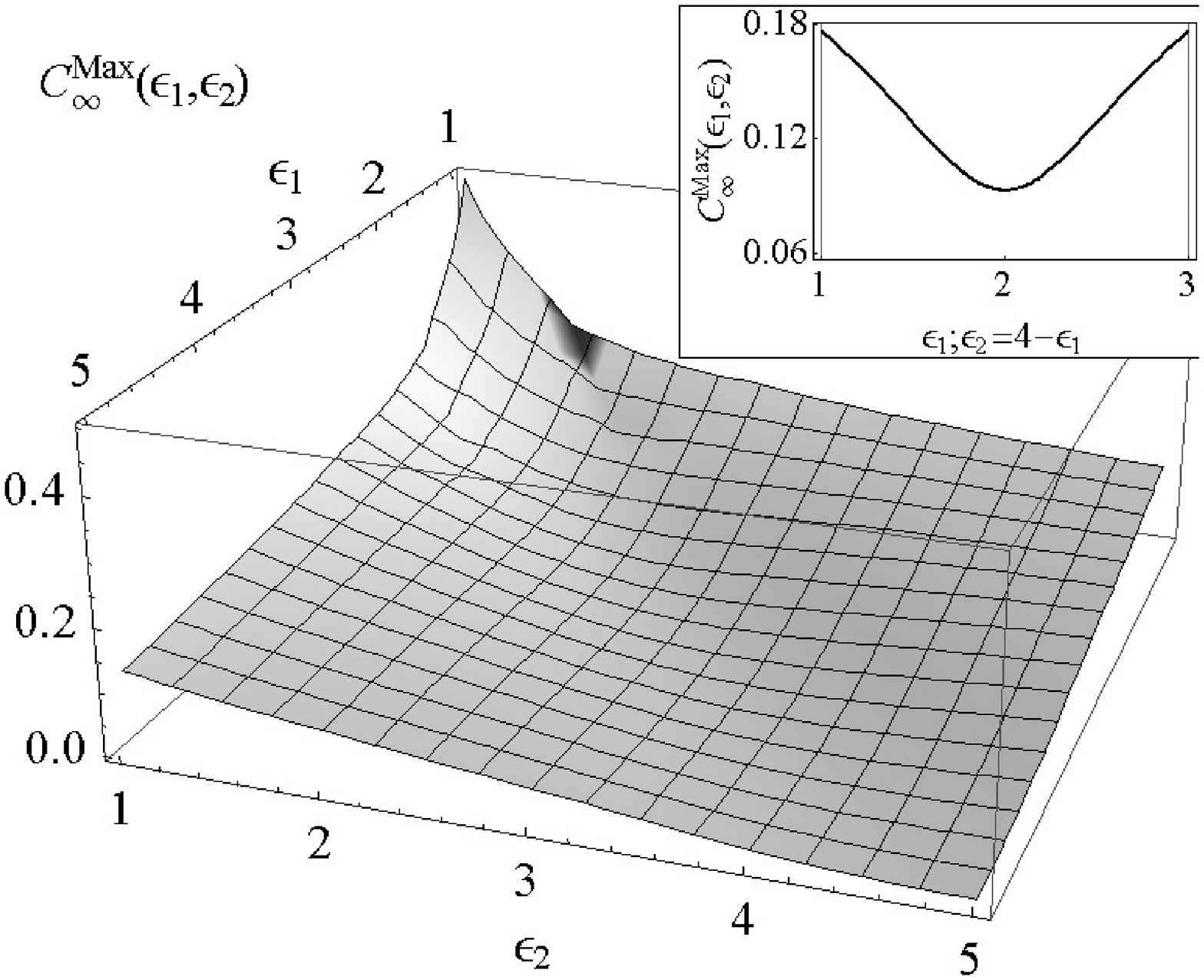}
\caption{Maximally possible steady-state concurrence
$C_{\infty}^{\mathrm{Max}}(\epsilon_1,\epsilon_2)$ in the weak
coupling case $\epsilon_1,\epsilon_2>K$ for $K=1$; In the corner:
profile of the 3D surface at the line $\epsilon_1+\epsilon_2=4$.}
\end{figure}

\section{RESULTS AND DISCUSSION}

The dynamics of entanglement is analyzed in Figures 1 and 2. In
Figure 1 the dynamics of the concurrence between the two qubits is
shown. For the model considered here the spin chain Hamiltonian $\hat{H}_S$ can entangle
the qubits for specific times, which gives rise to the
oscillations of concurrence one observes for short times (note
that the initial state is chosen to be separable
$\hat{\rho}_0=|1,0\rangle\langle1,0|$). For large times, the
system converges to its steady-state. One can see the
disappearance of entanglement with increasing temperatures of the
bosonic baths which was shown for the steady-state in \cite{Qui}.
In Figure 2 the dynamics of the concurrence for different initial
states of the qubits in shown. For all cases
the system converges to one and the same value of
entanglement. The plots in Figures 1 and 2 show clearly the
competition between unitary and dissipative dynamics. If the
qubits starts from a ``symmetric state", i.e.
$|1,1\rangle$, no oscillations in the concurrence
dynamics are observed and if the qubits start from a
``non-symmetric state", i.e. $|1,0\rangle\langle1,0|$ or
$(|1,0\rangle-|0,1\rangle)/\sqrt{2}$,
one can see oscillations of the concurrence which correspond to
the energy exchange between the qubits in the unitary evolution.
Both figures reveal that after time of order $t\sim2/\gamma$ the
concurrence ``forgets" about initial conditions and converges to
the same value, given by $C_\infty$ from the end of the Sec. III.
The steady-state concurrence
$C_\infty(\epsilon_1,\epsilon_2,K,T_1,T_2)$ is analyzed in
Figures 3, 4, 5 and 6. In Figures 3 and 4 we plot the
steady-state concurrence for the symmetric and
non-symmetric cases as a function of the mean temperature
($T_M=(T_1+T_2)/2$) and the temperature difference ($\Delta
T=T_1-T_2$) of the baths.
In the symmetric case one can easily see that the maximal value
of the entanglement is reached for equal bath temperatures ($\Delta
T=0$)
 $$C_{\mathrm{sym}}^{\mathrm{eq}}=\frac{\sinh(1/T)-1}{2\cosh(\omega_1/2T)\cosh(\omega_2/2T)}.$$
The critical temperature in units of $K$ above which the steady-state becomes separable is given by $T_C=\operatorname{arcsinh}(1)^{-1}$
($T_C\approx1.136$) \cite{Qui}.
It is interesting to note that in the non-symmetric case (Figure
4) the maximal entanglement is reached in the
non-equilibrium case ($\Delta T\neq0$). In particular, the maximal
entanglement is larger  than the corresponding non-symmetric
equilibrium concurrence
$$C_{\mathrm{non-sym}}^{\mathrm{eq}}=\frac{\sin\theta\sinh((\omega_2-\omega_1)/2T)-1}{2\cosh(\omega_1/2T)\cosh(\omega_2/2T)}.$$
The temperature at which entanglement disappears is a function
of the energy difference $\Delta\epsilon$ between qubits:
$$T_C=\frac{\sqrt{\Delta\epsilon^2/4+1}}{\operatorname{arcsinh}\sqrt{\Delta\epsilon^2/4+1}}.$$
It is easy to see that
this function reaches its minimum value in the symmetric case ($\Delta\epsilon=0$).
In Figures 5 and 6 we show the maximally reachable value of
entanglement as a function of qubits energies in the strong and
weak coupling case. For every pair of energies
$(\epsilon_1,\epsilon_2)$ we maximize the value of the concurrence
for the different temperatures of the baths ($T_1,T_2$). One can
see that in the strong coupling case ($\epsilon_1,\epsilon_2<K$;
Figure 5) the maximal value of the entanglement corresponds to the
symmetric case. In Figure 6 one can see that in
the weak coupling case ($\epsilon_1,\epsilon_2>K$) the maximal
value of the entanglement is reached in the non-symmetric case.

In conclusion, we have found an analytical solution for a simple
spin system coupled to bosonic baths at different temperatures.
We studied the dynamics of the system and showed that on the long
term the system converges to the steady-state solution. Resolving the entanglement dynamics allowed us to distinguish
between entanglement created by the system and by the bath.  For the
symmetric case ($\epsilon_1=\epsilon_2$) we
reproduced the steady-state found by \cite{Qui}.
We focused on the non-symmetric case
($\epsilon_1\neq\epsilon_2$) where we found that the steady-state
concurrence assumes its maximal value for unequal bath temperatures.
This is corresponds to a dynamical equilibrium, where the spin chain
transfers heat between the baths. We also found that a difference in local energy levels can make the steady-state
entanglement more stable against high temperatures. These
analytical results motivate further numerical studies on longer spin chains.

\begin{acknowledgments}
This work is based upon research supported by the South African
Research Chair Initiative of the Department of Science and
Technology and National Research Foundation.

DB would like to thank FP for the hospitality at UKZN and
acknowledges support by the QIP-IRC and Wolfson College, Oxford.
\end{acknowledgments}

\end{document}